\documentclass[12pt,dviwindows]{article}

\usepackage{color,epsf,graphicx}
\usepackage{amsmath,amssymb}
\usepackage{psfrag}
\usepackage[colorlinks=true,citecolor=blue,linkcolor=red]{hyperref}

\renewcommand{\bar}[1]{\overline{#1}}
\begin{document}

\begin{center}
{\Large
\bf Decay Rate Ratios of $\Upsilon (5S)\to B{\bar{B}}$ Reactions}\\

%\vspace{13pt}

\vspace{1.5cm}

Dae Sung Hwang and Hyungsuk Son\\
{\it{Department of Physics, Sejong University, Seoul 143--747,
Korea}}\\

\vspace{2.5cm}

{\bf Abstract}\\
\end{center}
\noindent
We calculate the decay rate ratios for OZI allowed decays
of $\Upsilon (5S)$ to two $B$ mesons
by using the decay amplitudes which incorporate the wave function of
the $\Upsilon (5S)$ state.
We obtain the results that the branching ratio of the $\Upsilon (5S)$ decay
to $B_s^*{\bar{B}}_s^*$ is much larger than the branching ratio to
$B_s{\bar{B}}_s^{*}$ or ${\bar{B}}_sB_s^{*}$,
in good agreement with recent experimental results of CLEO and BELLE.
This agreement with the experimental results is made possible
since the nodes of the $\Upsilon (5S)$ radial wave function induce
the nodes of the decay amplitude.
We find that the results for the $\Upsilon (5S)$ decays to
$B_u^{(*)}{\bar{B}}_u^{(*)}$ or $B_d^{(*)}{\bar{B}}_d^{(*)}$ pairs are
dependent on the parameter values used for the potential between heavy quarks.
\\

\vfill

\noindent
PACS codes: 12.39.-x, 12.39.Pn, 13.25.-k, 13.25.Gv\\
Key words: Quarkonium, B Meson, Hadronic Decay, Node of Amplitude

%\noindent
%$^a$e-mail: dshwang@sejong.ac.kr\\
\thispagestyle{empty}
\pagebreak

\setlength{\baselineskip}{13pt}

\section{Introduction}

The CLEO Collaboration recently observed the $B_s$ mesons in the $e^+e^-$
annihilation at the $\Upsilon (5S)$ resonance \cite{cleo06a}.
They established that $B_s$ meson production proceeds dominantly through the
creation of $B_s^*{\bar{B}}_s^*$ pairs.
They found $\sigma (e^+e^-\to B_s^*{\bar{B}}_s^*)=
[0.11^{+0.04}_{-0.03}({\rm stat})\pm 0.02({\rm syst})]\ {\rm nb}$
and set the following limits \cite{cleo06a}:
\begin{eqnarray}
{\sigma (e^+e^-\to B_s{\bar{B}}_s)\over \sigma (e^+e^-\to B_s^*{\bar{B}}_s^*)}
\qquad\qquad\ \ \ \
&<& 0.16 \ ,
\label{up1}\\
{[\sigma (e^+e^-\to B_s{\bar{B}}_s^*)\ +\
\sigma (e^+e^-\to B_s^*{\bar{B}}_s)]
\over \sigma (e^+e^-\to B_s^*{\bar{B}}_s^*)}
&<& 0.16 \ .
\label{up2}
\end{eqnarray}
The BELLE Collaboration collected data at the $\Upsilon (5S)$ resonance and
obtained the following ratio at the $\Upsilon (5S)$ energy \cite{belle07}:
\begin{equation}
{\sigma (e^+e^-\to B_s^*{\bar{B}}_s^*)\over
\sigma (e^+e^-\to B_s^{(*)}{\bar{B}}_s^{(*)})}
\ =\ 93^{+7}_{-9}\pm 1\ \%\ .
\label{up2belle}
\end{equation}
The above results of CLEO and BELLE show that
the branching ratio of the $\Upsilon (5S)$ decay to $B_s^*{\bar{B}}_s^*$ is
much larger than others ones to $B_s^{(*)}{\bar{B}}_s^{(*)}$ pairs,
and in particular that the ratio of
$[\sigma (e^+e^-\to B_s{\bar{B}}_s^*)\ +\
\sigma (e^+e^-\to B_s^*{\bar{B}}_s)]$ to
$\sigma (e^+e^-\to B_s^*{\bar{B}}_s^*)$ is much smaller
than the value which is expected from the spin countings.
These results imply that there exists a delicate mechanism in the decays of
$\Upsilon (5S)\to B{\bar{B}}$.

The unitarized quark model \cite{uqm} could predict similar results for the
above ratios.
These ratios were also calculated in Ref. \cite{BH87}
by using the decay amplitudes which take into account the bound state effect
through the wave function of the $\Upsilon (5S)$ state.
Even though these theoretical calculations showed the nature of the above experimental
results, they were performed before the masses of the
$B$ mesons, especially $B_s$ and $B_s^*$, are measured.
Therefore, it is desirable to perform a theoretical calculation by using
the experimentally established mass values of all $B$ mesons in order to obtain
the results which are more reliable quantitatively.
It is the purpose of this paper.
We use the same method for the calculation as that used in Ref. \cite{BH87}
with the experimentally measured values of $B$ meson masses.
We perform the numerical calculations for two sets of the parameter values for
the potential between heavy quarks, in order to see the dependence of the results on
the parameter values.

The CLEO Collaboration also measured the cross sections of the $\Upsilon (5S)$
decays to $B^*{\bar{B}}^*$, $B{\bar{B}}^*$, and $B{\bar{B}}$,
and found that the decay to $B^*{\bar{B}}^*$ is dominant with $(74\pm 15)$\% of
the total $B$ rate \cite{cleo06b}.
Here, $B=B_u$ or $B_d$, and $B{\bar{B}}^*$ signifies both
$B{\bar{B}}^*$ and $B^*{\bar{B}}$.
They summarized the measured branching ratios as follows \cite{cleo06a,cleo06b,cleo07}:
\begin{eqnarray}
{\cal{B}}(B^*{\bar{B}}^*)&=&43.6\pm 8.3\pm 7.2\ \%\ ,
\nonumber\\
{\cal{B}}(B{\bar{B}}^*+B^*{\bar{B}})&=&14.3\pm 5.3\pm 2.7\ \%\ ,
\nonumber\\
{\cal{B}}(B{\bar{B}})&<&13.8\ \%\ ,
\label{up3}\\
{\cal{B}}(B_s^{(*)}{\bar{B}}_s^{(*)})&=&16.8\pm 2.6^{+6.7}_{-3.4}\ \%\ \
({\rm using}\ D_s \ {\rm yields})\ ,
\nonumber\\
{\cal{B}}(B_s^{(*)}{\bar{B}}_s^{(*)})&=&24.6\pm 2.9^{+11.0}_{-5.3}\ \%\ \
({\rm using}\ \phi \ {\rm yields})\ .
\nonumber
\end{eqnarray}
The above branching ratios show that ${\cal{B}}(B^*{\bar{B}}^*)$ is
about two or three times of ${\cal{B}}(B_s^{(*)}{\bar{B}}_s^{(*)})$.
In this paper we also compare the results of our calculations with the above
experimental results.

%Their measured values of the production rates are given by \cite{cleo06b}
%\begin{eqnarray}
%\sigma (\Upsilon (5S)\to B^*{\bar{B}}^*)&=&(0.131\pm 0.025\pm 0.014)\ {\rm nb}\ ,
%\nonumber\\
%\sigma (\Upsilon (5S)\to B{\bar{B}}^*)\ &=&(0.043\pm 0.016\pm 0.006)\ {\rm nb}\ ,
%\label{up3}\\
%\sigma (\Upsilon (5S)\to B{\bar{B}})\ \ &<& \, \, 0.038\ {\rm nb}\ .
%\nonumber
%\end{eqnarray}

This paper is organized as follows. In section 2 we briefly summarize the
mechanism of light quark pair creation and the formula of the decay amplitude.
%Even though it is explained in detail in Refs. \cite{eichten12,zam1,byers94},
%we think that a concise summary is useful for this paper.
In section 3 we explain the procedures and results of our calculations.
The last section is conclusion.

%\section{Mechanism of Light Quark Pair Creation}
\section{Formula for Decay Amplitude of $\Upsilon (5S)\to B{\bar{B}}$}

The Cornell group studied the effect of OZI allowed decay channels \cite{eichten12}.
They proposed that the following interaction hamiltonian is responsible for the decay
as well as the binding of quark--antiquark bound states \cite{eichten12}.
\begin{equation}
H_I = {1\over 2} \sum_{k=1}^8 \int d^3x d^3y : \rho_k({\bf x})
V({\bf x}-{\bf y})\rho_k({\bf y}) : ,
\label{a13}
\end{equation}
where
\begin{equation}
V(r)=-{\kappa\over r}+{r\over a^2}.
\label{a14}
\end{equation}
In (\ref{a13}) $\rho_k({\bf x})=\psi^{\dagger}({\bf x}){1\over 2}\lambda_k\psi({\bf x})$
are the color densities of quark fields.
This model corresponds to the vector coupling since (\ref{a13}) is the leading term of
the vector coupling hamiltonian in the nonrelativistic expansion.

The decay rate of $\Upsilon (5S)$ with spin $s$ (which is 1) and mass $M$
to $B$ mesons with spin $s_1$ and $s_2$ is given by the formula \cite{BH87}
\begin{equation}
\Gamma = {1\over 8\pi} {P\over M} E_1 E_2 C(s s_1 s_2) |A_{5S}(P)|^2\ ,
\label{e1}
\end{equation}
where $P$ is the magnitude of the center of mass momentum of the mesons
and $E_1$ and $E_2$ are their energies.
The spin counting factor $C(s s_1 s_2)$ are for $s=1$ given be
${1/3}$, ${4/3}$, and ${7/3}$, for the cases
$s_1=s_2=0$, $s_1=0$ $s_2=1$ + $s_1=1$ $s_2=0$, and $s_1=s_2=1$, respectively
\cite{eichten12}.

\begin{figure}
\centering
\psfrag{xperp}[cc][cc]{$x_\perp$}
%\vspace*{0.5cm}
\begin{minipage}[t]{12.5cm}
\centering
\includegraphics[width=\textwidth]{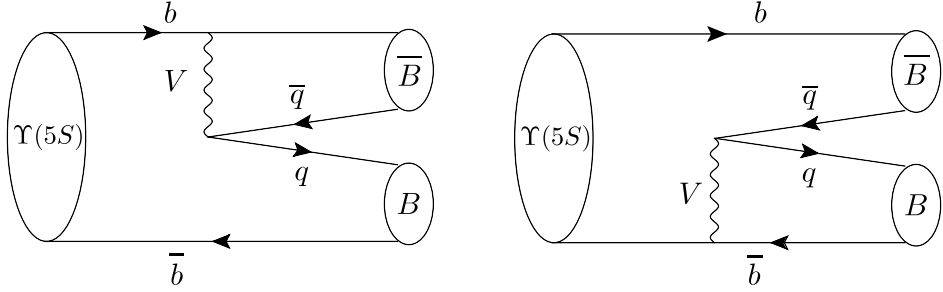}
%(a1)
\end{minipage}\hspace{0.0cm}
\vspace*{1.0cm}
%%%%
%%%%
\parbox{0.95\textwidth}{\caption{
Diagrams used to calculate the decay amplitude $A_{5S}(P)$.
\label{Diagrams1}}}
\end{figure}

%\vspace*{-4.0cm}

\begin{figure}
\centering
\psfrag{xperp}[cc][cc]{$x_\perp$}
%\vspace*{0.5cm}
\begin{minipage}[t]{7.0cm}
\centering
\includegraphics[width=\textwidth]{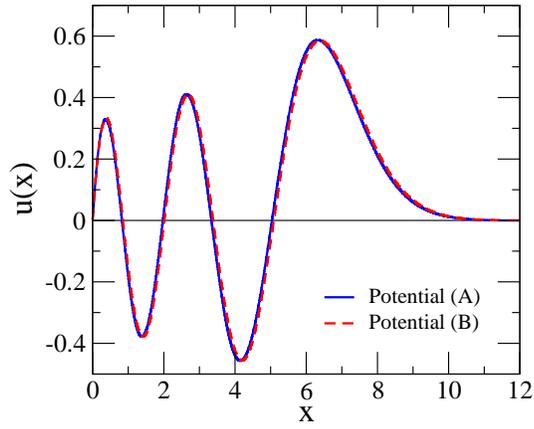}
%(a1)
\end{minipage}\hspace{0.0cm}
\vspace*{1.0cm}
%%%%
%%%%
\parbox{0.95\textwidth}{\caption{
$u(x)$ which is related to the radial wave function
%given from the radial wave function $R_{50}(r)$ of $\Upsilon (5S)$
%by the relation 
by $R_{50}(r)=( {m_b / a^2} {)}^{1/2}{u(x) / x}$, where
the dimensionless variable $x$ given by 
$x=( {m_b / a^2} {)}^{1/3}r$.
\label{comparison5}}}
\end{figure}

The decay amplitude $A_{5S}(P)$ calculated from the diagrams in Fig. \ref{Diagrams1}
is given by  \cite{eichten12,zam1,byers94,ELQ}
\begin{equation}
A_{5S}(P)=4\pi f_q I_{nL}^l(P)\ ,\qquad
f_q=\Big( {3\over \pi}{\Big)}^{1\over 2}\ \Big( m_q {m_bm_q\over m_b+m_q}{\Big)}^{-1}
\ ;\qquad q=u,d,s\ ,
\label{e2}
\end{equation}
where for the vector coupling of (\ref{a13})
the momentum dependent function $I_{nL}^l(P)$ is given by
\cite{eichten12,zam1,byers94,ELQ}
\begin{equation}
I_{nL}^l(P)=\int_0^{\infty}dt \Theta (t) R_{nL} ({t\over {\sqrt{\beta}}})
j_l({\rho_b P t \over {\sqrt{\beta}}})
\label{a18}
\end{equation}
with
\begin{equation}
\Theta (t)=[te^{-t^2}+(t^2-1)e^{-t^2/2}{\sqrt{\pi\over 2}} erf({t\over {\sqrt{2}}})]
+4\beta a^2 \kappa [-te^{-t^2}+e^{-t^2/2}{\sqrt{\pi\over 2}} erf({t\over {\sqrt{2}}})] ,
\label{a19}
\end{equation}
where
$R_{nL}(r)$ is the radial wave function of the heavy quark-antiquark bound state,
$j_l({\rho_b P t \over {\sqrt{\beta}}})$ is the spherical Bessel function,
and $\rho_b = m_b/(m_b+m_q)$.
We note that $I_{nL}^l(P)$ incorporates the radial wave function $ R_{nL}(r)$ of
the $\Upsilon (5S)$ state.
The first and second terms in (\ref{a19}) come from the linear and Coulombic parts
of (\ref{a14}), respectively.
Following \cite{eichten12},
we use the radial wave functions of the final $c{\bar{u}}$ and $c{\bar{d}}$
states with $L=0$ which are given
by the ground state harmonic-oscillator wave functions
$(4\pi)^{1\over 2}({2\beta\over\pi})^{3\over 4}{\rm exp}(-\beta r^2)$.
For the value of $\beta$ in the calculations of (\ref{a18}) and (\ref{a19}),
we use $\beta =({1\over 2a^2})({4\mu a\over 3{\sqrt{\pi}}})^{2\over 3}$
\cite{eichten12}, where $\mu$ is the reduced mass $m_q\rho_b$
of the $c{\bar{u}}$, $c{\bar{d}}$ or $c{\bar{s}}$ system.

%\vspace{1.5cm}

\begin{figure}
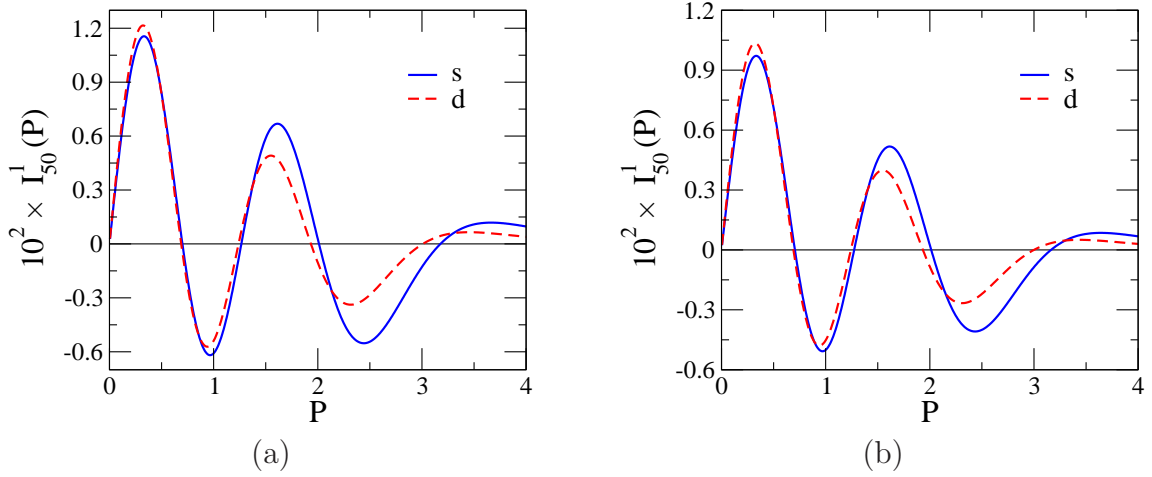

\centering
\psfrag{xperp}[cc][cc]{$x_\perp$}
%\vspace*{0.5cm}
\begin{minipage}[t]{7.0cm}
\centering
\includegraphics[width=\textwidth]{InL_A_LC.eps}
(a)
\end{minipage}\hspace{1.0cm}
%\vspace{2.0cm}
%%%%
\begin{minipage}[t]{7.0cm}
\centering
\includegraphics[width=\textwidth]{InL_A_L.eps}
(b)
\end{minipage}\hspace{0.0cm}
%%%%
\parbox{0.95\textwidth}{\caption{
$I_{50}^1(P)$ for the potential (A).
The solid line is for the $B_s^{(*)}$ meson pairs with $m_s =0.55$ GeV,
and the dotted line is for the $B_u^{(*)}$ or $B_d^{(*)}$ meson pairs
with $m_u =m_d =0.33$ GeV.
\label{comparison5a}}}
\end{figure}

\begin{figure}
\centering
\psfrag{xperp}[cc][cc]{$x_\perp$}
%\vspace*{0.5cm}
\begin{minipage}[t]{7.0cm}
\centering
\includegraphics[width=\textwidth]{InL_B_LC.eps}
(a)
\end{minipage}\hspace{1.0cm}
%\vspace{2.0cm}
%%%%
\begin{minipage}[t]{7.0cm}
\centering
\includegraphics[width=\textwidth]{InL_B_L.eps}
(b)
\end{minipage}\hspace{0.0cm}
%%%%
\parbox{0.95\textwidth}{\caption{
$I_{50}^1(P)$ for the potential (B).
The solid line is for the $B_s^{(*)}$ meson pairs with $m_s =0.55$ GeV,
and the dotted line is for the $B_u^{(*)}$ or $B_d^{(*)}$ meson pairs
with $m_u =m_d =0.33$ GeV.
\label{comparison5b}}}
\end{figure}

\begin{figure}
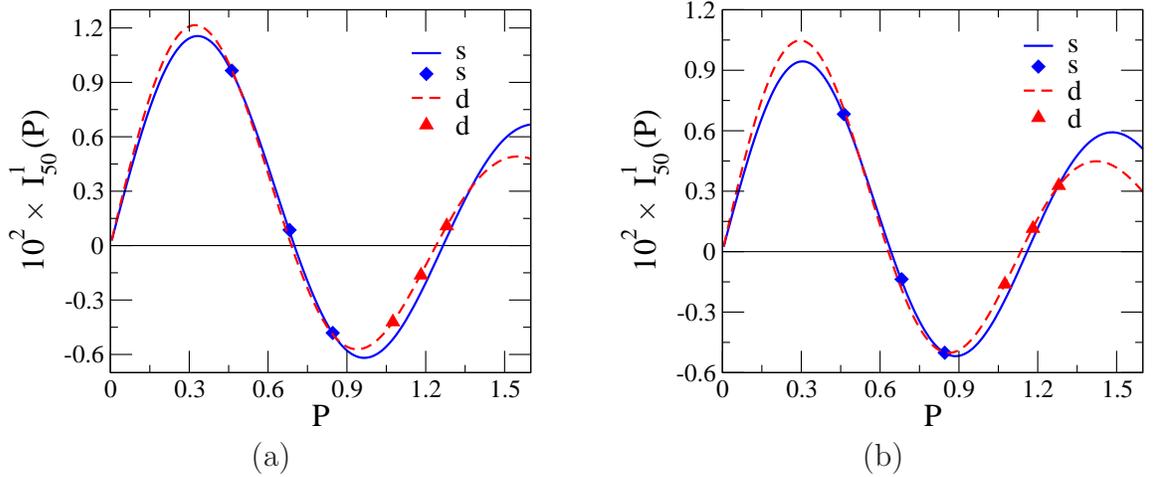

\centering
\psfrag{xperp}[cc][cc]{$x_\perp$}
%\vspace*{0.5cm}
\begin{minipage}[t]{7.0cm}
\centering
\includegraphics[width=\textwidth]{Position_InL_A.eps}
(a)
\end{minipage}\hspace{1.0cm}
%\vspace{2.0cm}
%%%%
\begin{minipage}[t]{7.0cm}
\centering
\includegraphics[width=\textwidth]{Position_InL_B.eps}
(b)
\end{minipage}\hspace{0.0cm}
%%%%
\parbox{0.95\textwidth}{\caption{
The position of the final momentum $P$ of the $B$ meson for each mode
in the $\Upsilon (5S)$ rest frame.
(a) is for the potential (A) and (b) is for the potential (B).
\label{comparison5c}}}
\end{figure}

\begingroup
%\squeezetable
\begin{table}[t]
%\vspace*{0.3cm}
\label{tab:mesonmass}
\begin{center}
\begin{tabular}{ccccccc}
\hline
$\Upsilon (5S)$&$B_u$&$B_d$&$B_u^*$&$B_d^*$&
$B_s$&$B_s^*$\\
\hline
10865&5279.15&5279.53&5325.1&5325.1&
5366.3&5412.8\\
\hline
\end{tabular}
\end{center}
\vspace*{-0.5cm}
\caption{Meson Masses (MeV) used in the calculation \cite{pdg08}.}
\end{table}
\endgroup

\begingroup
%\squeezetable
\begin{table}[t]
%\vspace*{0.3cm}
\label{tab:mesonmass}
\begin{center}
\begin{tabular}{ccccccccc}
\hline
$B_u{\bar{B}}_u$&$B_u{\bar{B}}_u^*$&$B_u^*{\bar{B}}_u^*$&
$B_d{\bar{B}}_d$&$B_d{\bar{B}}_d^*$&$B_d^*{\bar{B}}_d^*$&
$B_s{\bar{B}}_s$&$B_s{\bar{B}}_s^*$&$B_s^*{\bar{B}}_s^*$\\
\hline
1.282&1.183&1.075&1.280&1.182&1.075&0.846&0.682&0.462\\
\hline
\end{tabular}
\end{center}
\vspace*{-0.5cm}
\caption{Decay momenta in GeV for $\Upsilon (5S)$ calculated
with masses given in Table 1.}
\end{table}
\endgroup

\begingroup
%\squeezetable
\begin{table}[t]
%\vspace*{0.3cm}
\label{tab:mesonmass}
\begin{center}
\begin{tabular}{cccccccc}
\hline\hline
&Potential&$B_s^*{\bar{B}}_s^*$&$B_s{\bar{B}}_s^*+{\bar{B}}_sB_s^*$&$B_s{\bar{B}}_s$\ \ \ &
$B_d^*{\bar{B}}_d^*$&$B_d{\bar{B}}_d^*+{\bar{B}}_dB_d^*$&$B_d{\bar{B}}_d$\\
\hline\hline
(A)& L + C&1&0.007&\ \ 0.065\ \ \ \ \ \ &3.159&0.298&0.036\\
\hline
(A)& L only&1&0.010&\ \ 0.059\ \ \ \ \ \ &3.142&0.312&0.030\\
\hline\hline
(B)& L + C&1&0.034&\ \ 0.141\ \ \ \ \ \ &0.923&0.294&0.646\\
\hline
(B)& L only&1&0.026&\ \ 0.130\ \ \ \ \ \ &0.953&0.242&0.579\\
\hline\hline
\end{tabular}
\end{center}
\vspace*{-0.5cm}
\caption{Decay rate ratios of $\Upsilon (5S)\to B{\bar{B}}$ modes.
Potentials (A) and (B) are given in the text. ``L + C'' and ``L only'' mean that
we used ``both linear and Coulomb parts'' and ``only linear part'' respectively for the
calculation of (\ref{a19}). We always used both linear and Coulomb parts for the
calculation of the wave function of $\Upsilon (5S)$.}
\end{table}
\endgroup

\section{Results of Calculation}

We calculate the decay rates for OZI allowed decays
of $\Upsilon (5S)$ to two $B$ mesons
by using the formula (\ref{e1}) with the decay amplitudes given in (\ref{e2})
and (\ref{a18}).
For $V(r)$ in (\ref{a14}) we use the following two parameterizations in our
calculation:\\
(A) Eichten et al. \cite{eichten12}:
$\kappa$=0.517, $a$=2.12 ${\rm GeV}^{-1}$, $m_b$=5.17 GeV.\\
(B) Hagiwara et al. \cite{hagiwara}:
$\kappa$=0.47, $a=1/{\sqrt{0.19}}$ ${\rm GeV}^{-1}$, $m_b$=4.75 GeV.\\
For the masses of $u,d$ and $s$ quarks, we use
$m_{u,d}$=0.33 GeV and $m_s$=0.55 GeV.

The radial wave functions of $\Upsilon (5S)$ obtained by
solving the Schr\"dinger equation with the parameter sets (A) and (B)
are presented in Fig. 2.
The function $I_{50}^l(P)$ in (\ref{a18}) obtained with the above (A) and (B) sets of
the parameter values are
presented in Figs. \ref{comparison5a} and \ref{comparison5b}.

In order to obtain the decay amplitude for each final state of $B$ meson pair
from (\ref{e2}) and (\ref{a18}), we need to know the final momentum $P$ of the
$B$ meson in the $\Upsilon (5S)$ rest frame.
%$A_{5S}(P)$
We use the meson masses presented in Table 1,
which give the value $P$ for each mode as written in Table 2.
In Fig. \ref{comparison5c} we marked the position of the value $P$ for each mode,
from which we can get the magnitude of $I_{50}^l(P)$ for each mode of $B$ meson pair.
Then, we can calculate the decay rate $\Gamma$ from the formula (\ref{e1}) for
each mode of the $B$ meson pair.
Using the values of $\Gamma$ obtained in this way,
we calculate the ratios of the decay rates. 
We present our results of this calculation in Table 3.
When we compare our results presented in Table 3 with the experimental results
in (\ref{up3}),
for the $B_d^{(*)}{\bar{B}}_d^{(*)}$ pairs we should use two times the values
of Table 3 since in (\ref{up3}) $B$ can be either neutral or charged
\cite{cleo06b,cleo07}.
In our calculation the results for the $B_u^{(*)}{\bar{B}}_u^{(*)}$ pairs are
the same as those for the $B_d^{(*)}{\bar{B}}_d^{(*)}$ pairs
since we use the same value $m_{u,d}$=0.33 GeV for the $u$ and $d$ quark masses.

We find in Table 3 that the results
for the $B_s^{(*)}{\bar{B}}_s^{(*)}$ pairs
are in good agreement with the experimental results
in both cases of (A) and (B) sets of the parameter values.
Especially, we note that the result for the ratio
$[\sigma (e^+e^-\to B_s{\bar{B}}_s^*)\ +\
\sigma (e^+e^-\to B_s^*{\bar{B}}_s)]/\sigma (e^+e^-\to B_s^*{\bar{B}}_s^*)$
is consistent with the experimentally obtained very small value.
The reason why we could obtain such a consistent result is that we incorporate
the radial wave function of the $\Upsilon (5S)$ state which has nodes.
The nodes of the radial wave function induce
the nodes of the decay amplitude as we can see
in Figs. \ref{comparison5a} and \ref{comparison5b}.

We find that the results for the $\Upsilon (5S)$ decays to
$B_u^{(*)}{\bar{B}}_u^{(*)}$ or $B_d^{(*)}{\bar{B}}_d^{(*)}$ pairs are
sensitive to the parameter values used for the potential between heavy quarks.
For example, the result with the set (A) for the ratio
$[{\cal{B}}(B_u^{*}{\bar{B}}_u^{*})+{\cal{B}}(B_d^{*}{\bar{B}}_d^{*})]
/{\cal{B}}(B_s^{*}{\bar{B}}_s^{*})$
is rather larger than the experimental result and
the result with the set (B) for this ratio is 
within the error bar of the experimental result.
Such a situation happens since the node positions of the decay amplitude
are sensitive to the parameter values of the potential
in the range of large final $B$ meson momentum.

\section{Conclusion}

The CLEO and BELLE Collaborations recently performed experiments at the
$\Upsilon (5S)$ resonance.
Their results for the ratio
$[\sigma (e^+e^-\to B_s{\bar{B}}_s^*)\ +\
\sigma (e^+e^-\to B_s^*{\bar{B}}_s)]/\sigma (e^+e^-\to B_s^*{\bar{B}}_s^*)$
at the $\Upsilon (5S)$ energy is much smaller than the value expected from
the spin countings.
This delicate phenomenon was predicted by the unitarized quark model \cite{uqm}
and by the decay amplitudes which incorporate the wave function of
the $\Upsilon (5S)$ state \cite{BH87,eichten12}.
However, these theoretical studies were performed before the masses of the
$B$ mesons, especially $B_s$ and $B_s^*$, are known.
We use the same method for the calculation as that in Ref. \cite{BH87}
with the experimentally measured values of $B$ meson masses.
We perform the numerical calculations for two sets of the parameter values for
the potential between heavy quarks, in order to see the dependence of the results on
the parameter values used.

Our results for the ratios for the $B_s^{(*)}{\bar{B}}_s^{(*)}$ pairs are
in good agreement in both cases of (A) and (B) sets of the parameter values
of the potential between heavy quarks.
It is especially important that the result of our calculation agrees well with
the delicate experimental 
result that the branching ratio to $B_s{\bar{B}}_s^{*}$ or ${\bar{B}}_sB_s^{*}$
pair is much smaller than that to $B_s^*{\bar{B}}_s^*$ pair.
Obtaining such results by our calculation was possible since we take into consideration
the radial wave function of the $\Upsilon (5S)$ state which has nodes,
and the nodes of the radial wave function induce
the nodes of the decay amplitude of the $\Upsilon (5S)\to B{\bar{B}}$ reaction.
However, our results for the $\Upsilon (5S)$ decays to
$B_u^{(*)}{\bar{B}}_u^{(*)}$ or $B_d^{(*)}{\bar{B}}_d^{(*)}$ pairs are
sensitive to the parameter values used for the potential between heavy quarks
since the node positions of the decay amplitude are sensitive to the parameter values
in the range of large final $B$ meson momentum.

%%%%%%%%%%%%%%%%%%%%%%%%%%%%%%%%%%%%%%
\section*{Acknowledgments}
%%%%%%%%%%%%%%%%%%%%%%%%%%%%%%%%%%%%%%
One of the authors (D.S.H.) wishes to thank Nina Byers
for helpful discussions.
The authors wish to thank Jonghyun Kim for useful discussions.
This work was supported in part by the International Cooperation
Program of the KICOS (Korea Foundation for International Cooperation
of Science \& Technology),
and in part by the Korea Research Foundation Grant funded by the Korean
Government
%(MOEHRD, Basic Reserach Promotion Fund)
(KRF-2008-313-C00166).

\end{document}